\documentclass[12pt]{iopart}

\renewcommand{\vec}[1]{{\bi{#1}}}
\newlength{\overlinelength}                      
\renewcommand{\bar}[1]{\settowidth{\overlinelength}{#1}%
  \hspace{0.125\overlinelength}%
  \overline{\hspace{-0.125\overlinelength}#1%
    \hspace{-0.125\overlinelength}  }%
  \hspace{0.125\overlinelength}}

\usepackage[dvips]{epsfig}
\usepackage{iopams}
\usepackage{amstext}

\begin{document}
%\centerline{\fbox{\small PREPRINT: \hspace{0.5cm} Date: 04/23/1999 \hspace{0.5cm} Rev: 1.3
%  \hspace{0.5cm} Status: to be resubmitted \hspace{0.5cm}}}

\jl{3}

\title[Influence of spin-flip scattering \dots in a two-band Hubbard model]
      {Influence of spin-flip scattering on the stability of
        ferromagnetism in a two-band Hubbard model} 

\author{D Meyer and W Nolting}
\address{Humboldt-Universit\"at zu Berlin, Institut f\"ur Physik,               
           Invalidenstr.\ 110, 10115 Berlin, Germany}
\date{\today}

%%%%%%%%%%%%%%%%%%%%%%%%%%%%%%%%%%%%%%%%%%%%%%%%%%%%%%%%%%%%%%%%%%%%%%

\begin{abstract}
We investigate the influence of an interband exchange interaction on 
magnetism in a two-band Hubbard model. Our main emphasis lies on
spin-flip scattering which is often neglected but is neccessary to
retain the full rotational symmetry of the Hamiltonian. 
We find a striking dependence of the magnetization on the
interband exchange coupling constant $J$ and a substantial suppression of ferromagnetic order 
for a large range of values of $J$. The onset of an RKKY-like magnetic
ordering mechanism is also observed.
\end{abstract}
% 71.10.Fd       Lattice fermion models (Hubbard model, etc.)
% 71.20.Be       Transition metals and alloys
% 75.10.Lp       Magnetism: Band and itinerant models

%\submitted
%\maketitle
%
%
%
%%%%%%%%%%%%%%%%%%%%%%%%%%%%%%%%%%%%%%%%%%%%%%%%%%%%%%%%%%%%%%%%%%%%%%
%
%%%%%%%%%%%%%%%%%%%%%%    main text  %%%%%%%%%%%%%%%%%%%%%%%%%%%%%%%%
\section{Introduction}
Electronic correlations in the transition metals $Fe$, $Co$, $Ni$
still present a major challenge in condensed matter physics. The
stability of ferromagnetism in these materials is not yet fully
understood. Contrary to rare-earth systems, these materials have no
localized magnetic moments on ``atomic'' orbitals. The formation 
of finite magnetic moments in an itinerant electron system has to be
explained. 

In the past, there have been many attempts to describe transition
metals theoretically~\cite{OS84,NBDF89,BN90,SAS92,UIF94,NVF95,VN96}. Virtually 
all of these are based on a modification of the multi-band Hubbard
model~\cite{Hub63,Hub64a}, in which only the on-site Coulomb interaction
is considered. Recently, the importance of band degeneracy for correctly 
describing ferromagnetism has been confirmed by Quantum Monte Carlo (QMC)
calculations~\cite{HV98,Vea97}. But since all calculations mentioned
above rely on some approximations (even the quasi-exact QMC calculation
works only on a simplified Hamiltonian), no complete understanding of
the complicated $d$-band metals is reached yet.
The only way to gain a picture of the complex physics of these
systems is to consider partial problems for which conclusions can be
drawn from the available approximate approaches to more or less
oversimplified theoretical models. In this paper, we
want to provide another small piece belonging to the puzzle of the
$d$-band metals.

The investigation of ferromagnetism in the Hubbard model has a long
history. In fact, it was the original intention introducing this
model~\cite{Hub63,Kan63}.  It is known for a long time that for a band
filling of one electron above (above or below) half filling in the limit
$U\rightarrow\infty$, the ground state is ferromagnetic for a fcc- (sc-
or bcc-) lattice(Nagaoka
state)~\cite{Nag66,Tas99}. The stability of the Nagaoka state was subject
to extensive investigations. E.g. in infinite dimensions ($d=\infty$),
its stability could be proven for a wider range of band fillings and
finite $U$~\cite{Uhr96}. Using variational treatments, further limits
could be set on its stability on various lattice structures in two or
three dimensions~\cite{HUM97}. Related to these statements is the
so-called flat-band magnetism~\cite{Tas99,Mie91}. Here, a ferromagnetic
ground state could be rigorously proven for a dispersionless band
structure.  Furthermore, for finite temperatures, the existence of a
ferromagnetic phase for certain parameter ranges has been established
using dynamical mean-field theory~\cite{Ulm98,OPK97}.  We therefore
believe that the intraband Coulomb interaction as described by the
single-band Hubbard model is one major ingredient for itinerant
ferromagnetism.

However, besides the strong on-site Coulomb
interaction, transition metal systems are also characterized by the 5-fold
degeneracy of the $d$-bands. How does this fact influence ferromagnetism?
In atoms, Hund's rules will favor a parallel alignment of the spins of
electrons on degenerate levels. That orbital degeneracy
will enhance ferromagnetic stability also in lattice systems appears to
be a fact. The general validity of Hund's rule for the ground state of a degenerate
Hubbard model has indeed been proven~\cite{Mie93} (for further 
statements concerning ferromagnetism in degenerate Hubbard models see~\cite{Tas95,Tas99}).
Often, this is the justification for using a simplified interband
exchange interaction, which is restricted to a longitudinal Ising-like
spin exchange. The full SU(2)-symmetric interband exchange interaction
can be separated into a longitudinal (Ising) and a transversal
(spin-flip) term as will be shown below (see equation (\ref{hhj})).  The
longitudinal interaction will try to align the z-components of the spins
due to the energy gain connected with a positive value of interband
exchange constant $J$, thus fulfilling the predictions made by the
analogy to the atomic behaviour.  This is most obvious in mean-field 
theory, where the transverse part vanishes, and a
magnetization-dependent bandshift is induced by the longitudinal component.

In several recently published papers, more sophisticated calculations
for multi-band Hubbard models were presented, as e.g. in Quantum Monte
Carlo~\cite{HV98,HJC98}, slave-boson~\cite{FK97} or
Gutzwiller-variational ansatz methods~\cite{BGW97}. But many of these neglected
the transverse part of the spin exchange with the reasoning explained
above. Up to now it is not clear what the influences of the disregarded terms are.\\
A valuable contribution to the problem of ferromagnetism in orbitally
degenerate Hubbard models was given in~\cite{MK98}, where the authors
use an exact diagonalization method on a restricted Hilbert space of an 
in the limit of $d=\infty$ equivalent impurity model.

The aim of this paper is to investigate the influence of spin-exchange
processes originating from the transverse part of the interband
interaction on ferromagnetic stability. We examine a minimal two-band model
which includes those parts of the general Coulomb interaction that
we believe to be the most important for the stability of
ferromagnetism~\cite{NBDF89,BN90,NVF95,VN96}. This leads to the following
Hamiltonian:
\begin{eqnarray}
  \label{modham}
  H=&H_0+H_U+H_J\\ \bs 
  &H_0=\sum_{\vec{k},m,\sigma} \epsilon_m(\vec{k}) \,a_{\vec{k},m,\sigma}^{\dagger}a_{\vec{k},m,\sigma}\\
  \label{hu}  &H_U= \frac{1}{2}U\sum_{i,m,\sigma} n_{i,m,\sigma}\,n_{i,m,-\sigma}\\ 
  \label{hj}  &H_J=-\frac{1}{2}J\sum_{i,m} \vec{\bsigma}_{i,m}\cdot\vec{\bsigma}_{i,\bar{m}}
\end{eqnarray}
We use the usual notation for the electron annihilation (creation)
operators $a_{\vec{k},m,\sigma}^{(\dagger)}$ with wavevector $\vec{k}$,
band index $m$ ($\bar{m}$ being the complementary band of the two-band
system) and spin $\sigma$. The free bands are described by the
dispersion $\epsilon_m(\vec{k})$, $U$ and $J$ are combinations of
the appropriate Coulomb matrix elements as introduced e.g. in
~\cite{NBDF89,NVF95,VN96}. The spin operators in $H_J$ are defined as
\begin{eqnarray}
  \sigma_{i,m}^{\sigma} = a_{i,m,\sigma}^{\dagger}a_{i,m,-\sigma} \\
  \sigma_{i,m}^z = \frac{1}{2} \sum_{\sigma} z_{\sigma}\,n_{i,m,\sigma}\nonumber
\end{eqnarray}
with $z_\sigma =+1$ ($-1$) for spin $\uparrow$ (spin
$\downarrow$). Within this Hubbard-type Hamiltonian, the intraband part
$H_U$ is able to produce ferromagnetism for sufficiently large $U$
~\cite{Ulm98,OPK97,HN97a}. The second interaction, $H_J$ introduces
interband exchange processes of two different kinds, as already
mentioned above. This can be seen in the following decomposition of
$H_J$
\begin{equation}
  \label{hhj}
  H_J=-\frac{1}{4}J\sum_{i,m,\sigma} \left(\sigma_{i,m}^{\sigma}
  \sigma_{i,\bar{m}}^{-\sigma}+\sigma_{i,m}^z\sigma_{i,\bar{m}}^z\right)
\end{equation}
The first term (transverse part) represents spin-flip scattering and the
second one (longitudinal) an ``Ising-like'' exchange, which tends to
stabilize spontaneous ferromagnetic order. Many calculations on
multi-band Hubbard models only consider the Ising term and neglect the
spin-flip part of the interaction~\cite{HV98,HJC98,FK97,BGW97}. In this paper, we want to
trace both parts of the interband interaction (\ref{hhj}) with identical quality.

It goes without saying that (\ref{hu}) and (\ref{hj}) do not respresent
the full set of local Coulomb interactions between $d$-electrons. Our
previous studies~\cite{NBDF89,NVF95,VN96,NHB92}, however, have evidenced 
that they are most important for treating magnetic phenomena in
transition metals. The final 
goal of our investigation will be the more or less quantitative 
description of real substances such as Fe~\cite{NVF95}, Co~\cite{VN96},
Ni~\cite{NBDF89}, and Gd~\cite{REN98}. For this purpose we combine
many-body model methods with ``\textit{ab-initio}''-bandstructure
calculations. By definition the underlying many-body model incorporates
only those interactions which are believed to be decisive for the
collective magnetism with respect to temperature dependencies and
typical correlation effects, and which are probably not properly taken
into account by usual LSDA-treatments. According to our previous
calculations the interactions (\ref{hu}) and (\ref{hj}) should be 
most important while, e.g., the interband Coulomb interaction (denoted 
by $\bar{U}$ in our previous studies~\cite{NVF95}) turns out to be not
that decisive for the band-ferromagnetism of transition metals. It can
be assumed that this part of the local Coulomb interaction is
sufficiently well covered by an LSDA calculation. In this sense our
present model investigation is to be understood. Instead of tackling a
more complicated, but probably still insufficient Hamiltonian, we
aim at a better
understanding of those interactions which might be the roots of the
phenomenon \textit{ferromagnetism}, without referring to a real
system. The description of more realistic systems seems possible by a
combination of our model study with LDA calculations along the kines of
ref.~\cite{NBDF89,NVF95,VN96}, which is intended
for the future.

In the following section, we introduce an \textit{effective medium
method} to decompose the complicated many-body problem of the
Hamiltonian (\ref{modham}) into two separately solvable problems of
simpler structure. After that we introduce the approximations that lead
us to a fully self-consistent solution for the two-band model. In
section \ref{cha:results}, we present and discuss our results.

\section{Theory}
\subsection{Effective Medium Approach}
Even though we restricted ourselves to a rather simple model
Hamiltonian, one is in need of a convincing approximation method to
solve the problem. We propose a method based on an effective medium
ansatz. This ansatz will map the original problem of Hamiltonian
(\ref{modham}) onto a set of simpler model Hamiltonians, for which well 
tested standard approximations exist. These can be put together in an
appropriate way to get a solution of the original problem.
This method is generalizable to many different models, a similar
approach applied to the periodic Anderson model was recently published~\cite{MNRR98}.

We want to introduce the method using the two-band Hamiltonian
(\ref{modham}) with the two interaction terms $H_U$ and
$H_J$. The self-energy $\Sigma_{\vec{k},m,\sigma}(E)$ can be defined using the equation of motion
of the single-electron Green's function
$G_{\vec{k},m,\sigma}(E)=\langle\!\langle a_{\vec{k},m,\sigma}; a_{\vec{k},m,\sigma}^{\dagger}\rangle\!\rangle$:
\begin{eqnarray}
  \label{eq:eqmo}
        E G_{\vec{k},m,\sigma}(E) &= \hbar + \epsilon_m(\vec{k})
  G_{\vec{k},m,\sigma}(E) +
  \langle\!\langle
  [a_{\vec{k},m,\sigma},H_U+H_J]_-;a_{\vec{k},m,\sigma}^{\dagger}\rangle\!\rangle\nonumber\\
  &=\hbar + \epsilon_m(\vec{k})
  G_{\vec{k},m,\sigma}(E) + \Sigma_{\vec{k},m,\sigma}(E) G_{\vec{k},m,\sigma}(E)
\end{eqnarray}
Using the linearity of the
commutator and the Green's function, one can define ``self-energy
parts'',
\begin{eqnarray}
\langle\!\langle[a_{\vec{k},m,\sigma},H_U]_-;a_{\vec{k},m,\sigma}^{\dagger}\rangle\!\rangle 
  = \Sigma_{\vec{k},m,\sigma}^{(U)}(E) G_{\vec{k},m,\sigma}(E)\\
  \langle\!\langle[a_{\vec{k},m,\sigma},H_J]_-;a_{\vec{k},m,\sigma}^{\dagger}\rangle\!\rangle 
  = \Sigma_{\vec{k},m,\sigma}^{(J)}(E) G_{\vec{k},m,\sigma}(E)
\end{eqnarray}
with
$\Sigma_{\vec{k},m,\sigma}(E)=\Sigma_{\vec{k},m,\sigma}^{(U)}(E)+\Sigma_{\vec{k},m,\sigma}^{(J)}(E)$.
Assuming the knowledge of either one of these self-energy parts, one
can introduce the following effective Hamiltonians:
\begin{eqnarray}
  \label{heff}
  H_{\text{eff}}^{(U,\eta)}= \sum_{\vec{k}} \left(\epsilon_m(\vec{k}) +
  \Sigma_{\vec{k},m,\sigma}^{(J)}(\eta)\right)
  a_{\vec{k},m,\sigma}^{\dagger}a_{\vec{k},m,\sigma} + H_U\\
  \label{heff2}
  H_{\text{eff}}^{(J,\eta)}= \sum_{\vec{k}} \left(\epsilon_m(\vec{k}) +
  \Sigma_{\vec{k},m,\sigma}^{(U)}(\eta)\right)
  a_{\vec{k},m,\sigma}^{\dagger}a_{\vec{k},m,\sigma} + H_J
\end{eqnarray}
which formally depend on a parameter $\eta$. By solving each of these
Hamiltonians for all values of $\eta$, one can obtain the respectively
missing self-energy part using the following identities:
\begin{eqnarray}
\label{se}
  \Sigma_{\vec{k},m,\sigma}^{(U)}(E)=\Sigma_{\vec{k},m,\sigma}^{(\text{eff},U,\eta)}(E)|_{\eta=E}\\
  \label{se2}
  \Sigma_{\vec{k},m,\sigma}^{(J)}(E)=\Sigma_{\vec{k},m,\sigma}^{(\text{eff},J,\eta)}(E)|_{\eta=E}
\end{eqnarray}
with $\Sigma_{\vec{k},m,\sigma}^{(\text{eff},U,\eta)}(E)$ being the
self-energy of the effective Hamiltonian (\ref{heff}) and
$\Sigma_{\vec{k},m,\sigma}^{(\text{eff},J,\eta)}(E)$ of Hamiltonian
(\ref{heff2}). The introduction of the energy-parameter $\eta$ in the
effective Hamiltonians (\ref{heff}) and (\ref{heff2}) is neccessary in
order to distinguish two different kinds of energy. The effective
medium energy parameter $\eta$ should not be confused with the energy as 
dynamic variable of e.g. an
equation of motion method to solve the respective partial many-body
problem. This distinction is mandatory since otherwise, the system could
show unphysical behaviour. Furthermore, for the same reasoning it is
neccessary to calculate expectation values in the full system, e.g. by
using the formal solution of equation (\ref{eq:eqmo}) for the single-electron 
Green's function:
\begin{equation}
  G_{\vec{k},m,\sigma}(E)=\frac{\hbar}{E-\left(\epsilon_m(\vec{k})+\Sigma_{\vec{k},m,\sigma}^{(U)}(E)+\Sigma_{\vec{k},m,\sigma}^{(J)}(E)\right)}
\end{equation}
The two problems posed by the Hamiltonians (\ref{heff}) and
(\ref{heff2}) are strongly related to each other. The solution of one in 
form of the respective self-energy (\ref{se}) or (\ref{se2}) is needed
as input for the other. This implies a self-consistency condition
on the two self-energy parts which can only be fullfilled in an iterative
way. But the advantage of the effective medium
ansatz is also rather obvious: The two Hamiltonians (\ref{heff}) and
(\ref{heff2}) have been analyzed very well by now since they are both
standard models of many-body-theory. Model (\ref{heff}) is essentially
the single-band Hubbard model~\cite{Hub63}, and model (\ref{heff2}) is
known as $sf$-model or ferromagnetic Kondo-lattice~\cite{Nag74,Nol79}.  So for
both partial problems already known approximation schemes can be used.

For the Hubbard-model, many useful approximations exist. We use the
spectral density approximation (SDA)~\cite{HN97a,Nol72,NB89}, which has
to be considered as a strong-coupling theory. Although this choice
prohibits an investigation of the small-$U$ behaviour, its advantages
are enormous. Besides of being mathematical simple and numerically
reasonably fast, it has proven to give a qualitatively correct picture of 
ferromagnetism in the strong-coupling regime and compares to more
sophisticated approaches~\cite{PHWN98}. It will be described in the following
chapter. The $sf$-model has also
attracted much interest. A very promising interpolating,
moment-conserving equation-of-motion decoupling scheme has been
developed in refs. ~\cite{NDB93,NMR96,NRM97}. This method will be
discussed in section \ref{rsatheo}. The combination of these two
calculations along the lines described above as effective medium
approach will lead us to the solution of the full problem posed by
Hamiltonian (\ref{modham}).

\subsection{Spectral density approach}
\label{sdatheo}
According to the effective medium ansatz, the self-energy part
$\Sigma_{\vec{k},m,\sigma}^{(U)}(E)$ connected with the intraband
Hubbard interaction will be calculated using the Hamiltonian
(\ref{heff}). We use the spectral density approach (SDA) to solve this
problem. This method is numerically simple and fast, and, as shown in
the limit of infinite dimensions, its magnetic properties resemble
qualitatively the quasi-exact Quantum Monte Carlo
calculation~\cite{Ulm98,PHWN98,HN97b}. The SDA has been studied for
the single-band Hubbard model extensively~\cite{HN97a,Nol72,NB89}. In the
following we will give only a short outline of the
calculations. Starting point is the single-electron spectral density
being defined by
\begin{equation}
  \label{spec_dens}
\fl  S_{\vec{k},m,\sigma}(E)= \frac{1}{N} \sum_{i,j} \exp (i \vec{k}
    (\vec{R}_i-\vec{R}_j))\frac{1}{2\pi}\,\int\limits_{-\infty}^{+\infty}dE\,
  \exp(-\frac{i}{\hbar}Et) \, \langle
  [a_{i,m,\sigma}(t),a_{j,m,\sigma}^{\dagger}(0)]_+\rangle \nonumber
\end{equation}
where $[\dots,\dots]_+$ denotes the anticommutator and $\langle\dots\rangle$ the
thermodynamic average. The construction operators are taken to be in the
Heisenberg time-dependent picture.

In an exact spectral-moment analysis in the limit $U\rightarrow\infty$, Harris
and Lange have shown that the spectral density essentially
consists of a two-peak structure ~\cite{HL67}. Since ferromagnetism is
widely believed to be a strong coupling phenomenon, any reasonable
approximation aiming at ferromagnetism should contain this limiting
case~\cite{PHWN98}.

In the SDA, one makes the following ansatz for the spectral density,
which will turn out to correctly reproduce the positions and weights
of the quasiparticle peaks according to the Harris- and Lange-calculation in
the limit of $U\rightarrow\infty$. 
\begin{equation}
  S_{\vec{k},m,\sigma}(E)=\sum_{j=1,2}\hbar\,\alpha_{\vec{k},m,\sigma}^{(j)}\,\delta(E-E_{\vec{k},m,\sigma}^{(j)})
\end{equation}
The unknown parameters  $E_{\vec{k},m,\sigma}^{(j)}$ and $\alpha_{\vec{k},m,\sigma}^{(j)}$, the
quasiparticle energy and spectral weight can be calculated by the moment 
method. That means they are fitted by the use of the
first four moments of the spectral density, which represent several sum
rules, and which can be calculated directly from the Hamiltonian. 
\begin{eqnarray}
  \label{moments}
   M_{\vec{k},m,\sigma}^{(n)} = \int\limits_{-\infty}^{+\infty}dE\,E^n\,
    S_{\vec{k},m,\sigma}(E) \\
    =\langle  [ \underbrace{[\dots[a_{\vec{k},m,\sigma},H]_-,\dots,H]_-}_{\text{$n$-fold commutator}} ,
    a_{\vec{k},m,\sigma}^{\dagger} ]_+\rangle 
\end{eqnarray}
This procedure is identical to the one performed in ~\cite{HN97a} for
the conventional
Hubbard problem. An explicit description of the calculation is presented there.
As a result one obtains a self-energy of the following structure:
\begin{equation}
  \label{selfenergy_SDA}
  \fl \Sigma_{\vec{k},m,\sigma}^{(U)}(E)=U\,\langle n_{i,m,-\sigma}\rangle
   \frac{E-B_{m,-\sigma}-F_{\vec{k},m,-\sigma}}
  {E-B_{m,-\sigma}-F_{\vec{k},m,-\sigma}-U(1-\langle n_{i,m,-\sigma}\rangle)}
\end{equation}
The decisive terms are $B_{m,-\sigma}$ and $F_{\vec{k},m,-\sigma}$ which
distinguish this self-energy from the Hubbard-I solution ~\cite{Hub63}.
There these terms would be replaced simply by the center of gravity of
the appropriate Bloch band.  $B_{m,-\sigma}$ and $F_{\vec{k},m,-\sigma}$ 
mainly consist of higher correlation functions. They may provoke a
spin-dependent shift and/or deformation of the bands and may therefore
be responsible for the existence of spontaneous magnetism
~\cite{HN97a,PHWN98,PWN97}.
The $\vec{k}$-dependent term $F_{\vec{k},m,-\sigma}$ seems to be of
minor importance for the magnetic behaviour~\cite{HN97a}. Since
$\sum_{\vec{k}} F_{\vec{k},m,-\sigma}=0$ it does not change the center
of gravity of the density of states. It is mainly responsible
for a deformation and narrowing of the bands.  We have therefore
neglected this term in the following calculations. The term
$B_{m,-\sigma}$ has the following structure:
\begin{equation}
  \fl B_{m,\sigma}=\frac{1}{N}\sum_{i,j} \left(\sum_{\vec{k}}
  \exp(-i\vec{k}(\vec{R}_i-\vec{R}_j))
  \epsilon_{m,\sigma}(\vec{k})\right)
 \langle a_{i,m,\sigma}^{\dagger} a_{j,m,\sigma}(2\,n_{i,m,-\sigma}-1)\rangle
\end{equation}
with
$\epsilon_{m,\sigma}(\vec{k})=\epsilon_m(\vec{k})+\Sigma_{\vec{k},m,\sigma}^{(J)}(\eta)$.
Fortunately, this two-particle correlation function is accessible via the
single-electron spectral density, no higher Green's functions have to be calculated~\cite{GN88}.
One obtains the following expression:
\begin{eqnarray}
  \label{bandshift}
  \fl B_{m,\sigma}=\frac{1}{N}\sum_{\vec{k}}
  \epsilon_{m,\sigma}(\vec{k}) +\frac{1}{\langle
    n_{m,\sigma}\rangle(1-\langle n_{m,\sigma}\rangle)}   
   \frac{1}{N \, \hbar} \sum_{\vec{k}} \left(\epsilon_{m,\sigma}(\vec{k})-\frac{1}{N}\sum_{\vec{k'}}
    \epsilon_{m,\sigma}(\vec{k'})\right) \nonumber\\ 
  \times \int\limits_{-\infty}^{+\infty}d\widetilde{E}\, \frac{S_{\vec{k},m,\sigma}(\widetilde{E})}{e^{\beta\,(\widetilde{E}-\mu)}+1}
  \, \left(\frac{2}{U_m}\left(\widetilde{E}-\epsilon_{m,\sigma}(\vec{k})\right)-1\right)
\end{eqnarray}
This leads to a set of equations which can be solved self-consistently.
Despite its obvious restrictions, e.g. the complete neglect of
quasiparticle damping, the two-pole approximation together with the
moment method is able to describe
the magnetic properties of the Hubbard model surprisingly
well ~\cite{HN97a,HN97b}. Since the subject of this paper is the
influence of the spin-flip processes on the ferromagnetism introduced
by the on-site intraband Hubbard interaction, this choice of a
numerically simple procedure here seems reasonable.
However, one should bear in mind that conceptually, the SDA is a strong
coupling method which surely becomes questionable for intermediate to
weak couplings. Consequently, we restrict all the following
considerations to situations with $U_m$ substantially larger than the
free bandwidth $W_m$.

\subsection{Rigid Spin Approximation}
\label{rsatheo}
Next, we have to solve the Hamiltonian (\ref{heff2}).
The approximation scheme for this effective problem has to be
chosen very carefully since our investigation aims at effects directly
induced by the interaction (\ref{hj}). There is no standard
method to solve this 
model beyond mean-field level. In the following, we want to apply a
non-perturbative, moment-conserving, self-consistent method which 
explicitly includes spin exchange scattering.

The basis of the approximation scheme is the similarity between the
interaction (\ref{hj}) and the well-known $sf$- or Kondo-lattice
model. The difference of the two models lies in the electron spin
operator $\vec{\bsigma}_{i,\bar{m}}$. In our model, this operator is
built from electron construction operators. In the Kondo model, the
charge degrees of freedom of the $f$-spin have been projected out. Only
a pure spin operator remains. A
formal equivalency between both models can be reached if one
artificially fixes the operator $\vec{\bsigma}_{i,\bar{m}}$ to its spin
degrees of freedom. This can be done in an iterative way for both bands
separately. That means for calculating the self-energy for band $m$ one
has to fix the spin-operator of the other band ($\bsigma_{i,\bar{m}}$)
and vica versa. Thus no constraints are introduced concerning the
quantum mechanical attributes of the partial electron system under
consideration, as e.g. the non-distinguishability of the particles.  In
our opinion, the name \textit{Rigid Spin Approximation} (RSA) is an
appropriate denotation for this method.

Now we can apply a proper approximation developed for the $sf$-model. We use
a moment-conserving, self-consistent interpolating equation-of-motion
decoupling scheme~\cite{NDB93,NMR96,NRM97}. A discussion of the
neccessary approximations and their implications can be found there, so
that we can restrict ourselves to a short summary in this paper.
In the equation of motion for the single-electron
Green's function $G_{i,j,m,\sigma}(E)=\langle\!\langle a_{i,m,\sigma}; a_{j,m,\sigma}^{\dagger}\rangle\!\rangle$,
\begin{eqnarray}
    \fl E G_{i,j,m,\sigma}(E) =
    \hbar +
    \sum_l \left(\sum_{\vec{k}} \exp(i \vec{k}(\vec{R}_i-\vec{R}_l))
    (\epsilon_m(\vec{k})+\Sigma_{\vec{k},m,\sigma}^{(U)}(\eta))\right) 
    G_{l,j,m,\sigma}(E)\nonumber \\
    - \frac{1}{2}J\big(F_{i,i,j,m,\sigma}(E)+z_\sigma
    \Gamma_{i,i,j,m,\sigma}(E)\big)
\end{eqnarray}  
two higher Green's functions are introduced, the ``spin-flip'' function
$F_{i,l,j,m,\sigma}(E)=\langle\!\langle \sigma_{i,\bar{m}}^{-\sigma}
a_{l,m,-\sigma};a_{j,m,\sigma}^{\dagger}\rangle\!\rangle$ and the
``Ising'' function $\Gamma_{i,l,j,m,\sigma}(E)= \langle\!\langle
\sigma_{i,\bar{m}}^z
a_{l,m,\sigma};a_{j,m,\sigma}^{\dagger}\rangle\!\rangle$. For these two
Green's functions, the respective equations of motion can be obtained
without problems. In each of them, new Green's functions are
introduced. For these we carefully apply a sophisticated decoupling
scheme. The central idea is to express all local higher Green's function
in the equations of motion of the spin-flip and Ising function by
interpolation ``ansatzes'' between several non-trivial limiting cases,
such as $S=\frac{1}{2}$, ferromagnetic saturation of one band, empty and
full band. The corresponding interpolation parameters can be obtained
using a moment method similar to the one described in section
\ref{sdatheo}. Within this approximation scheme we obtain a selfenergy
of the following structure:
\begin{equation}
  \label{eq:rsa_sener}
    \Sigma_{m,\sigma}^{(J)}(E)=J\langle \sigma_{i,\bar{m}}^z\rangle
    + J^2 
    \mathcal{F}(\Sigma_{m,\sigma}^{(J)}(E),\langle
    n_{m,\sigma}\rangle, \langle
    \sigma_{i,\bar{m}}^+\sigma_{i,m}^-\rangle,\dots)
\end{equation}
The first term corresponds to the mean-field solution. The second part corresponds
to higher order terms in $J$. The complex functional $\mathcal{F}$ depends on several
correlation functions, as e.g. the interband spin-exchange correlation 
$\langle \sigma_{i,\bar{m}}^+\sigma_{i,m}^-\rangle$ or the interband
Ising correlation $\langle \sigma_{i,\bar{m}}^z\sigma_{i,m}^z\rangle$, the selfenergy
part $\Sigma_{m,\sigma}^{(J)}(E)$ and of course, via the effective
medium, on the  intraband selfenergy part
$\Sigma_{m,\sigma}^{(U)}(E)$. All correlation functions as well as the
selfenergy parts have to be determined selfconsistently. The selfenergy
part $\Sigma_{m,\sigma}^{(J)}(E)$ is $\vec{k}$-independent due to the
neglect of magnonic excitation energies, which are small compared to
the electronic excitations under consideration~\cite{NMR96}.
It is worth to mention that this method can be continued
smoothly to the exactly solvable non-trivial limiting case of one
electron in a ferromagnetic saturated background of $f$-spins.

\section{Results and Discussion}
\label{cha:results}
In this section we present the results obtained by the theory
described above. The results are compared with the usual mean-field
calculation for the spin-exchange interaction. The two-band model
under consideration consists of two bands $m=\{0,1\}$ of unit width
($W_{\{0,1\}}=1.0$) thus defining the energy unit used in this
paper. 
The bands are not degenerate, the centers of gravity are shifted by
$0.1$. As an example we choose tight-binding bcc free densities of
states~\cite{Jel69}. In the single-band Hubbard model, the existence of
ferromagnetism depends on the lattice structure, as has been shown by
various methods (stability of the Nagaoka state~\cite{HUM97},
SDA~\cite{HN97b}, QMC in $d=\infty$~\cite{Wea98}). From these
investigations, it follows that the system has stronger tendency
towards ferromagnetism in non-bipartite lattices (e.g. fcc). It seems
reasonable to believe that this will also hold for the two-band model
although we did no systematic analysis of the lattice dependence.

\begin{figure}[h]
  \begin{center}
    \epsfig{file=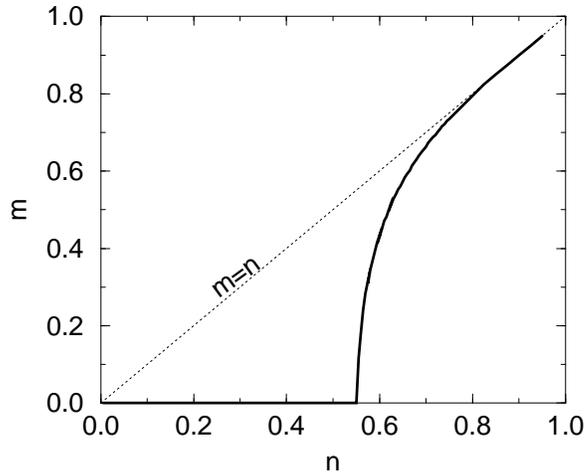,width=8cm}
    \caption{Magnetization $m$ as function of band occupation $n$
      of a single-band Hubbard model with $U=5$, $T=0$ on a bcc-lattice. The dashed line
      indicates saturation ($m=n$). Antiferromagnetic ordering is expected
      close to half filling ($n=1$) which is not considered in this
      paper. Due to particle-hole-symmetry, it is sufficient to
      investigate the region of $0<n<1$}
    \label{fig:hub_m_n}
  \end{center}
\end{figure}
One expects the local intraband Coulomb matrix
element $U$ to be large whereas the interband exchange coupling
constant $J$ could be one order of magnitude smaller. As mentioned above 
the SDA which we use to find the intraband selfenergy part
$\Sigma_{\vec{k},m,\sigma}^{(U)}(E)$ is basically a strong-coupling
theory. In the large-$U$ regime, however, magnetic key quantities such
as the Curie-temperature, the $T=0$-moment, and so on, are already
saturated, i.e. no longer $U$-dependent (see e.g. figure 11
in~\cite{HN97a}). More interesting is the $J$-dependence. We therefore
restrict ourselves to the representative value of $U=5$ and inspect in
detail the influence of the interband interaction $J$. Furthermore, the
evaluation is confined to $T=0$, although the theory of course holds for 
finite temperatures, too.

\subsection{The $J=0$ case}
In the case of vanishing interband coupling, $J=0$, the situation is
identical to two separate single-band Hubbard models which are only
coupled by a common Fermi energy. It determines the respective partial
band occupations $n_{\{0,1\}}$ according to the total number of electrons per site
$n_{tot}=n_0+n_1$.
For the single-band Hubbard model, the existence of
ferromagnetism has been since long a matter of controversial
discussions, but recent results confirm its stability for certain
parameter regimes~\cite{Uhr96,Ulm98}. The SDA has turned out to be able to reproduce the
QMC results in this limiting case on a qualitative level~\cite{PHWN98,HN97b}.

First, let us remind of a result obtained within a single-band
Hubbard model. In figure \ref{fig:hub_m_n}, the dependence of the
magnetization on the electron density (occupation number) is plotted.
For $n<n_c^{\text{(hub)}}\approx 0.56$ the system is paramagnetic, only for
$n>n_c^{\text{(hub)}}$ ferromagnetic ordering is possible. With increasing $n$,
the system becomes quickly saturated. Antiferromagnetic ordering occurs
only in the very vicinity of half filling ($n=1.0$) which we will never
consider in the following discussion.  More information on
ferromagnetism in the single-band Hubbard model can be found in
~\cite{Ulm98,OPK97,HN97a,NB89,HN97b,GN88,WPN98}.

\begin{figure}[htbp]
  \begin{center}
    \epsfig{file=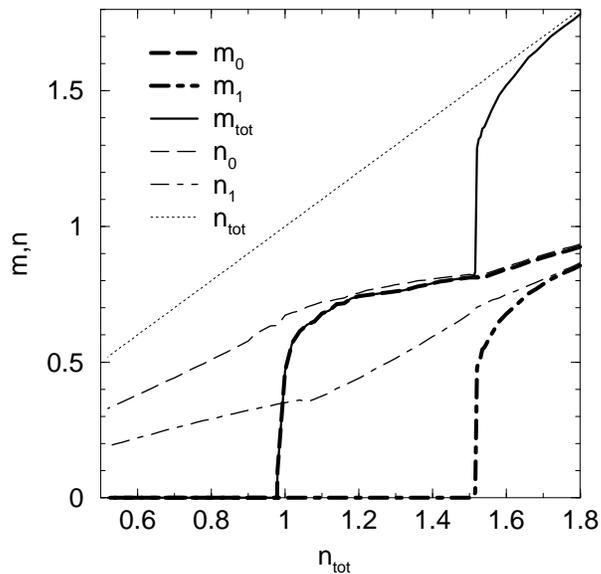,width=8cm}
    \caption{Magnetization $m$ as function of total occupation
      $n_{tot}$ of a two-band system (solid line: total magnetization,
      dashed line: polarization of the lower band ($m=0$), chained line:
      polarization of the upper band ($m=1$)). The thin lines represent the
      partial occupation numbers per band. System parameters: $U=5$, $J=0$, 
      $T=0$}
    \label{fig:m_n_j0}
  \end{center}
\end{figure}
In figure \ref{fig:m_n_j0}, the magnetization as function of the
electron density is plotted for the two-band situation. Together with
the total magnetization $m_{tot}$ the partial occupation numbers and
polarizations per band are shown. The interband interaction is still set
to zero ($J=0$).  As already mentioned above, both bands are coupled
only via the chemical potential. Ferromagnetic order sets in in band
$m=0$, the lower band, when this band reaches a critical occupation of
$n_0=n_{c,0}^{(J=0)}\approx 0.63>n_c^{\text{(hub)}}$ which is the case
for a total occupation of $n_{tot}\approx 0.98$. The critical electron
densisty of the lower band $n_{c,0}^{(J=0)}$ is larger than in the
single-band case ($n_c^{\text{(hub)}}$). Even though the upper band is
neither explicitely coupled to the lower band nor is ferromagnetic by
itself, it has an influence on the para- to ferromagnetic transition of
the lower band. This can be understood, since the upper band  can
act as source or sink of electrons for the lower band.
With the onset of ferromagnetism, a
spin-dependent band splitting takes place. This provides for a
rearrangement of electrons between the bands due to the same chemical
potential for both bands. Now, if the system tries to order
ferromagnetically, when band $m=0$ reaches the ``single-band critical
occupation'' of approximately $n_0\approx 0.56$, the corresponding shift
of the densities of states will make the ferromagnetic phase
instable. This happens until the real critical value of the two-band
situation, $n_0=n_{c,0}^{(J=0)}\approx 0.63$ is reached.  This effect is
analogous to the findings in ~\cite{BJ79}, where a similar situation was
examined using a Stoner-like theory.

\subsection{The $n_{tot}$-dependency of the magnetization}
\begin{figure}[h]
  \begin{center}
    \epsfig{file=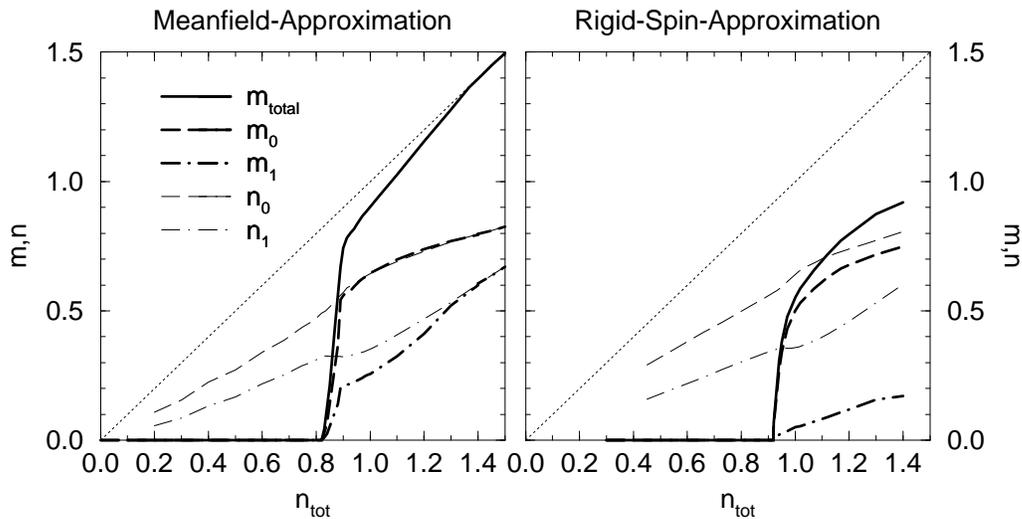,width=14cm}
    \caption{The same as figure \ref{fig:m_n_j0}, but with finite
      $J=0.2$. Left-hand side: Interband exchange interaction in
      mean-field approximation, right-hand side in RSA}
    \label{fig:m_n}
  \end{center}
\end{figure}

Let us focus on the interband-exchange interaction. Figure
\ref{fig:m_n} shows, similar to figure \ref{fig:m_n_j0}, the
magnetization as function of electron occupation but with finite
$J=0.2$. On the left hand
side, calculations were made using the mean-field approximation for
the interband-exchange self-energy part, on the right hand side the
rigid spin approximation as explained in chapter \ref{rsatheo} was applied.

The mean-field result is easily understood: When the lower band reaches
a critical occupation number $n_0=n_{c,0}^{\text{(mf)}}\approx
0.50<n_{c,0}^{(J=0)}$ a transition to a ferromagnetic state occurs. Now the
interband exchange in mean-field approximation will provide for a rigid
spin-dependent shift of the $m=1$ quasiparticle bands proportional
to the magnetization of the lower band. This induces a polarization of
the upper band as well. The same mechanism now works as a feedback onto the
lower band, increasing the magnetization even more. The lower band
becomes thus quickly saturated.

When analyzing the same situation using the rigid spin approximation,
the picture gets modified in a rather drastic way. The onset of
ferromagnetism is indicated by a critical occupation for the lower
band $n_c^{\text{(RSA)}}$ which is only a little larger than
$n_c^{\text{(mf)}}$. But
the polarization of the upper band is much weaker than in the mean-field
case. And furthermore, the lower band never reaches saturation. The
latter is not a result of the weak polarization of the $m=1$ band, but
both effects have the same origin. Continuous spin-flip scattering
prohibits the lower band from reaching saturation. These processes, in
addition to the generally stronger quasiparticle damping also reduce the
magnetic polarization of the upper band.

\subsection{The $J$-dependence of the magnetization}
\begin{figure}[htbp]
  \begin{center}
    \epsfig{file=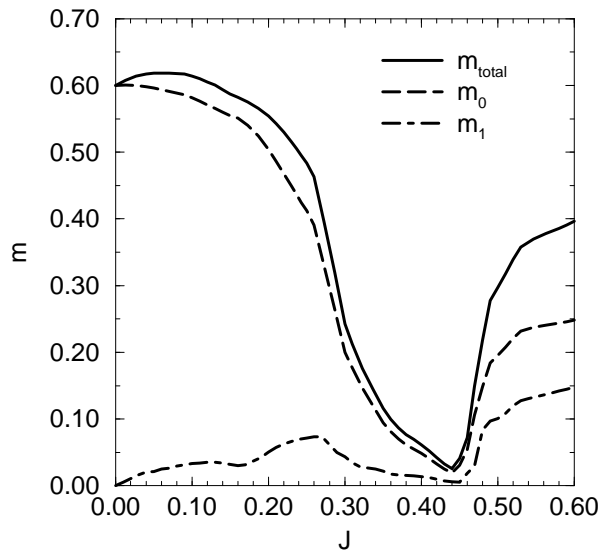,width=8cm}
    \caption{Total magnetization $m$ (solid line) and partial
      polarization of the two bands (dashed: lower band (0), chained: upper
      band (1)) as function of the interband coupling constant $J$ for fixed 
      total occupation $n_{tot}=1.0$ and $U=5$, $T=0$.}
    \label{fig:m_j}
  \end{center}
\end{figure}

\begin{figure}[htbp]
  \begin{center}
    \epsfig{file=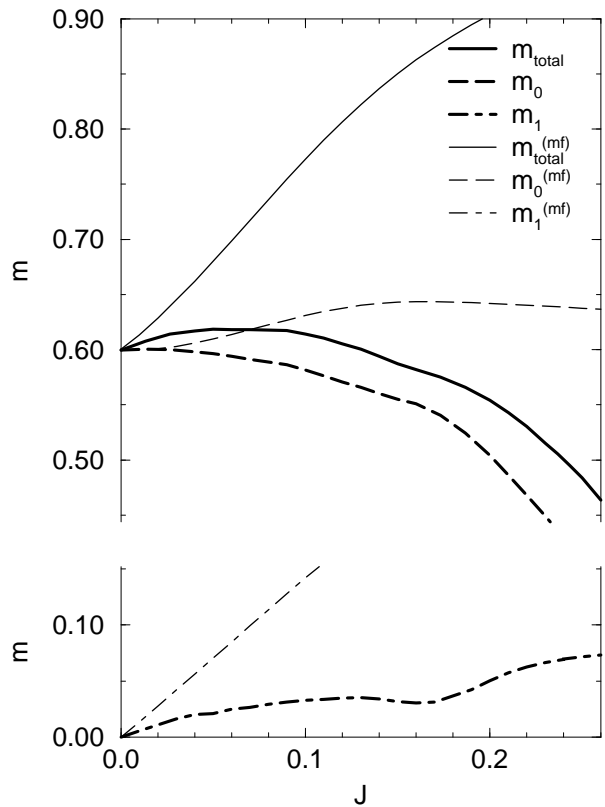,width=8cm}
    \caption{The same as figure 3, but zoomed to small values of
      $J$. Additionally, the results of the mean-field calculation are
      plotted (thin lines).}
    \label{fig:m_klein_j}
  \end{center}
\end{figure}
\begin{figure}[htbp]
  \begin{center}
    \epsfig{file=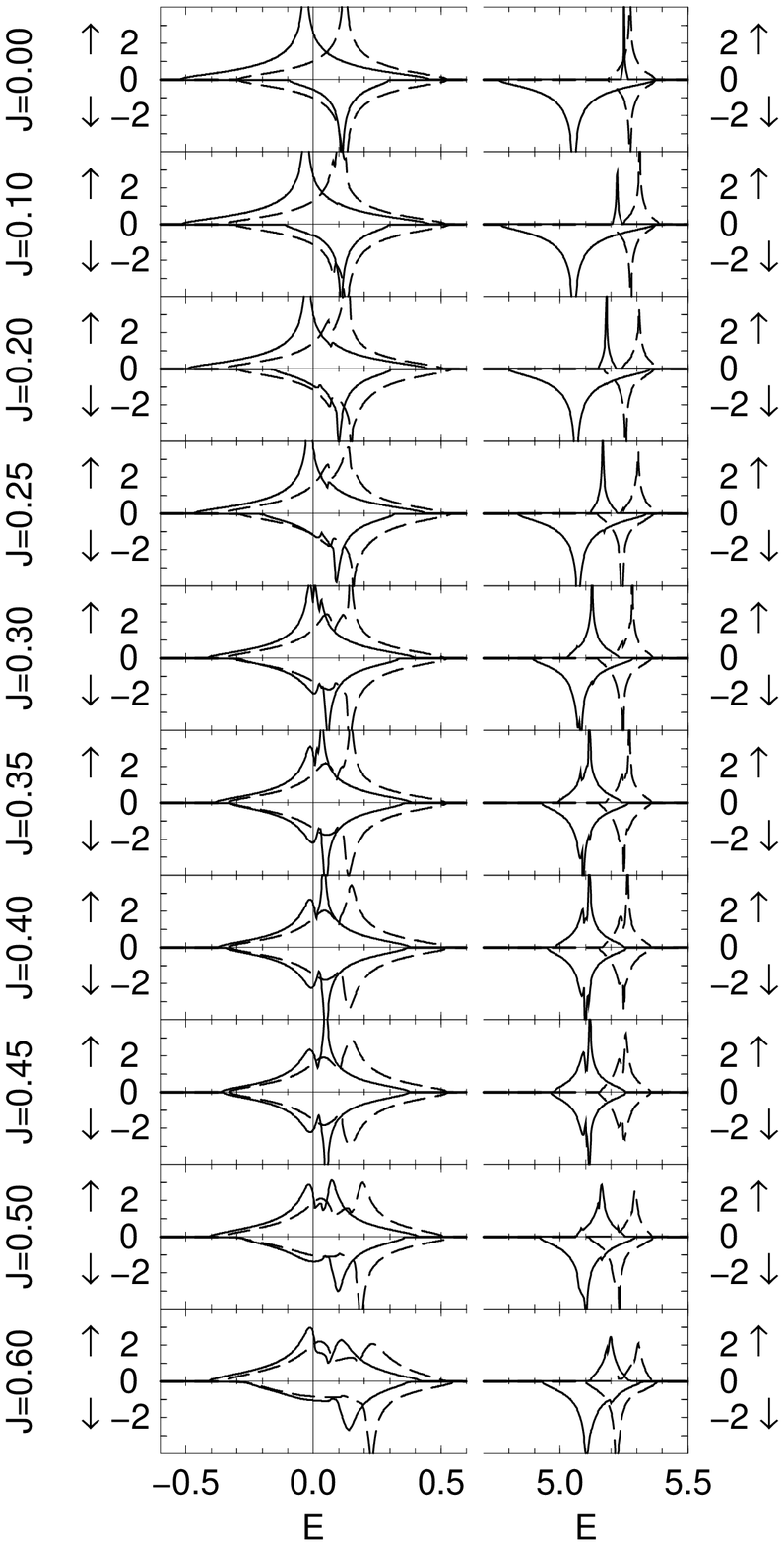,width=9cm}
    \caption{Quasiparticle densities of states (QDOS) for certain
      values of $J$ for the same system parameters as in figure 3 with the
      chemical potenial $\mu=0$. The $m=0$ bands are drawn in solid, the
      $m=1$ quasiparticle bands in dashed lines. Spin
      $\uparrow$ and spin $\downarrow$ QDOS are plotted separately as
      indicated.}
    \label{fig:qdos}
  \end{center}
\end{figure}
Next, we want to investigate the $J$-dependence of the
magnetization. This turns out to yield unexpected, non-trivial
results. Figure \ref{fig:m_j} shows the magnetization as function of $J$ 
for fixed $U=5$ and $n_{tot}=1.0$. In figure \ref{fig:m_klein_j}, the
behaviour for very small $J$ is plotted together with results obtained
using mean-field approximation for the interband-exchange, and finally, the 
corresponding quasiparticle densities of states are plotted in figure
\ref{fig:qdos}. For the chosen total occupation number $n_{tot}=1.0$,
the lower band is already ferromagnetically ordered for $J=0$, whereas
the upper band is still paramagnetic (see figure \ref{fig:m_n_j0}).

The magnetization curve in figure \ref{fig:m_j} can be separated 
roughly into 3 regions. For very small $J$ the magnetization rises
(region A), for
intermediate $J$ it decreases (region B) and shows a reentrant behaviour
for the largest $J$ under consideration (region C).

\subsubsection{Region A:}

The first of these regions is characterized by an increasing
magnetization as seen more clearly in figure
\ref{fig:m_klein_j}. There, additionally to the RSA result, the
mean-field curves are plotted. These are simply understood,
any finite $J$ will induce magnetic polarization into the upper band, and
via a feedback, push the lower band into saturation. Band $m=1$ will become
more and more polarized, finally reaching saturation, too (not plotted
in figure \ref{fig:m_klein_j}). The same effect can be seen in the RSA
calculation, which is in first order of $J$ identical to the mean-field
result. 

\subsubsection{Region B:}

Already for $J\approx 0.05$, which is still a very small parameter,
deviations from the mean-field result are quite strong. This leads to the
second regime in figure \ref{fig:m_j}. Here, the magnetization
decreases, first slightly, but with increasing $J$
stronger and stronger. In this parameter regime, the higher-order contributions
in $J$ to the self-energy become more and more
important. The connection between the reduction of the magnetization and
spin-flip processes becomes obvious by a comparison with the conventional
$sf$-model. The important observation is the onset of a small dip in the 
quasiparticle densities of states (QDOS) with
increasing $J$ (figure \ref{fig:qdos}).
This dip is on a much smaller
energy scale than the Hubbard splitting, which is also clearly visible
in figure \ref{fig:qdos}.

To understand this feature, we want to relate it to a feature known from
the conventional $sf$-model. There, a bandsplitting can occur for
intermediate to large values of $J$. The size of the gap scales roughly
with $J$. The physics causing this gap can be understood best by
examining the exactly solvable special case of one electron in a
saturated spin background~\cite{NolBd7,NMR96}. In this case two different
elementary excitations can be observed. One represents the scattering
of an electron accompanied by the
emission of a magnon, whereas the other can be connected with a bound
state of an electron with a cloud of magnons. The latter manifests
itself in the spectral density by a delta-like peak splitting of the
scattering part for large enough values of $J$. The respective
quasiparticle is called magnetic polaron~\cite{NMR96}.
For the general
case of the $sf$-model, i.e. for finite electron density and a not fully
polarized spin system, a similar band splitting due to the same two
elementary processes will occur for large enough
values of $J$ in both spin-resolved sub-bands~\cite{NRM97}.
The dip seen in the QDOS of figure \ref{fig:qdos} is a precursor
to the magnetic-polaron-induced bandsplitting. 

The reduction of the magnetization originates clearly from the spin-flip
terms in the Hamiltonian. A further test is the artificial neglect of
quasiparticle damping by setting $\Im \Sigma_{\sigma}(E)=0$. Even then,
a reduction, though smaller, of the magnetization with increasing $J$ is
found. So additionally to the generic damping effects, explicit
spin-flip scattering depolarizes the system.

\subsubsection{Region C:}

Finally, one observes a reentrant behaviour of the ferromagnetic
ordering in figure \ref{fig:m_j}, the origin of which is not completely
clear. One possible mechanism for supporting magnetic order would be a
RKKY-like interaction mediated by the interband interaction (\ref{hj}). 
The reentrant behaviour would indicate, that only for $J\gtrsim0.45$, the 
effective RKKY-interaction is strong enough to have an effect. To
support our proposal,
we have performed a calculation for the case of
$U=0$ and finite $J$. Even for these more or less unphysical parameters, 
we find stable ferromagnetic solutions. But as in the finite-$U$ case,
these only occur for 
relatively large values of $J\gtrsim0.64$. So even this artificially
castrated model without direkt Coulomb interaction shows similar
behaviour, which is, of course, quantitatively modified in the more
realistic model with finite $U$. \\
How this proposed mechanism is related to an enhanced stability of
ferromagnetism due to the two-band situation found in~\cite{MK98} is an
interesting, but open question.

\section{Conclusions}
In this paper, we have analyzed a special multi-band Hubbard model
which contains the two interaction terms believed to have the biggest
influence on the stability of ferromagnetic ordering.

Our model (\ref{modham}) is clearly insufficient to describe the
rich physics of transition metals. But the same applies probably to
any other model Hamiltonian which is still tractable within many-body
theory. However, \textit{ab-initio} calculations as e.g. density
functional theory applied within the local density approximation
(LDA) appear to underestimate just those correlation effects which
seem to be decisive for phenomena like ferromagnetism.
A proper combination of LDA
calculations with a many-body treatment gives a promising way to solve
these difficulties. The LDA calculation accounts for all interactions on
a mean-field level, the many-body treatment should restrict itself only
to the most important correlations. We believe that
the model (\ref{modham}) contains in this sense those interactions whose 
contributions beyond meanfield have the biggest impact on
magnetism. A combination of LDA calculations with the model
(\ref{modham}) using a simpler approximation on the exchange part yields 
very good agreement with experimental
facts~\cite{NBDF89,BN90,NVF95,VN96}. For example, by fitting the
interaction constants $U$ and $J$ to groundstate properties, the
Curie-temperature could be calculated astonishingly accurate.

In this paper, we introduced a more sophisticated approximation scheme,
which is especially improved on the exchange interaction part of model
(\ref{modham}). 
Our analysis was based on the fact that the intraband Hubbard
interaction alone is able to form ordered magnetic moments in a band
(itinerant magnetism). We investigated the influence of interband exchange
coupling, often referred to as Hund's rule coupling, on the stability 
of ferromagnetism. 
Only for very small values of $J$, the respective interband coupling constant,
we can verify the mean-field result which leads always to an enhanced stability of
spontaneous ferromagnetism. Already for $J\approx 0.1$
which we call intermediate coupling the magnetization gets suppressed
by spin-flip scattering (using the free bandwidth as energy scale:
$W=1$). With increasing $J$ the ferromagnetic order 
almost vanishes. The reentrant behaviour found for even larger $J$
might be due to an RKKY-like ordering mechanism mediated by the
interband interaction (\ref{hj}).
The regime of $J$ most often referred to transition
metals belongs to intermediate coupling strengths. Our results
indicate that the influence of the spin flip processes can manifest
itself in a rather dramatic reduction of the magnetization.

\appendix
\section*{References}
\bibliographystyle{/eibe/dmeyer/lit/uuu.pott}
\bibliography{/eibe/dmeyer/lit/artikel,/eibe/dmeyer/lit/artikel.nicht.vorhanden,/eibe/dmeyer/lit/buecher,/eibe/dmeyer/lit/arbeiten}
\ack
One of us (D.M.) gratefully acknowledges the support of the
\textit{Friedrich-Naumann foundation}. This work also benefitted from
the financial support of the \textit{Sonderforschungsbereich SFB 290}
(``Metallische d\"unne Schichten: Struktur, Magnetismus und elektronische 
Eigenschaften'') of the Deutsche Forschungsgemeinschaft.
\end{document}